\documentclass[reprint,aps,amsmath,amssymb,twoside,prd,showkeys,superscriptaddress,nofootinbib]{revtex4-2}
\usepackage[letterpaper,top=1.45cm,bottom=1.65cm,left=1.75cm,right=1.75cm]{geometry}
\pdfoutput=1
\usepackage{verbatim}
\usepackage{comment}
\usepackage{graphicx}
\usepackage[T1]{fontenc}
\usepackage{epsfig}
\usepackage{bm}
\usepackage{tensor}
\usepackage{amssymb}
\usepackage{float}
\usepackage{amsmath}
\usepackage{subfigure}
\usepackage{dcolumn}
\usepackage[utf8]{inputenc}
\usepackage{cancel}
\usepackage[colorlinks]{hyperref}
\usepackage[usenames,dvipsnames]{color}
\usepackage{lmodern}
\usepackage[x11names]{xcolor}
\usepackage{array, colortbl, hhline}
\usepackage{xparse}
\DeclareExpandableDocumentCommand\emptycell{O{|c|}m}{\multicolumn{#2}{#1}{}}

\usepackage{fancyhdr}
\fancypagestyle{plain}

\usepackage{placeins}

\hypersetup{
     breaklinks=true,
    pdfstartview={FitH},    
    colorlinks=true,       
    linkcolor=blue,          
    citecolor=red,        
    filecolor=magenta,      
    urlcolor=blue,           
    anchorcolor=green,      
    linktocpage=true
}

\allowdisplaybreaks

\begin{document}
\title{Well-posedness of the four-derivative scalar-tensor theory of gravity in singularity avoiding coordinates}

\author{Llibert Arest\'e Sal\'o}
\email{l.arestesalo@qmul.ac.uk}
\affiliation{School of Mathematical Sciences, Queen Mary University of London, Mile End Road, London, E1 4NS, United Kingdom}

\author{Katy Clough}
\email{k.clough@qmul.ac.uk}
\affiliation{School of Mathematical Sciences, Queen Mary University of London, Mile End Road, London, E1 4NS, United Kingdom}

\author{Pau Figueras}
\email{p.figueras@qmul.ac.uk}
\affiliation{School of Mathematical Sciences, Queen Mary University of London, Mile End Road, London, E1 4NS, United Kingdom}

\begin{abstract}
We show that the most general scalar-tensor theory of gravity up to four derivatives in $3+1$ dimensions is well-posed in a modified version of the CCZ4 formulation of the Einstein equations in singularity-avoiding coordinates. We demonstrate the robustness of our new formulation in practise by studying equal mass black hole binary mergers for different values of the coupling constants. Although our analysis of well-posedness is restricted to cases in which the couplings are small, we find that in simulations we are able to push the couplings to larger values, so that a certain weak coupling condition is order one, without instabilities developing. Our work provides the means for such simulations to be undertaken by the many numerical relativity codes that rely on the moving puncture gauge to evolve black hole singularities.
\end{abstract}

\maketitle
\thispagestyle{fancy}

\section{Introduction}
\label{sec:intro}

Detections of gravitational waves from the mergers of compact objects permit testing of the strong field, highly dynamical regime of general relativity (GR) \cite{LISA:2022kgy,Gnocchi:2019jzp,Barack:2018yly}. The current waveforms are tested for consistency with GR, mainly using methods that parameterise the deviations to the merger, inspiral and ringdown phases in a general way \cite{Krishnendu:2021fga, LIGOScientific:2021sio}. However, to check whether such parameterised deviations are well-motivated and consistent in alternative theories beyond GR, it is necessary to obtain predictions for specific models \cite{LISA:2022kgy, Okounkova:2022grv, Johnson-McDaniel:2021yge,Shiralilou:2021mfl,Perkins:2021mhb,Carson:2020ter,Carson:2020cqb}, with numerical relativity the essential tool for the merger section of the signal.

For a long time, a barrier to obtaining such predictions was that, for many modified gravity theories of interest, well-posed formulations were not known, and progress could only be made using order-reduced methods that potentially suffer from the accumulation of secular errors over long inspirals \cite{Okounkova:2022grv,Elley:2022ept,Doneva:2022byd,Okounkova:2020rqw,Silva:2020omi,Okounkova:2019zjf,Okounkova:2019dfo,Witek:2018dmd}. A great deal of effort has therefore been devoted to find mathematically well-posed formulations of certain alternative theories of gravity of interest. The property of well-posedness guarantees that, given some suitable initial data, the solution to the equations of motion exists, is unique and depends continuously on the initial data. Hence, it is an essential prerequisite to be able to simulate the theory on a computer and extract waveforms that can then be compared to the predictions of GR.  

Current observations indicate that the deviations from GR in the strong field regime are small \cite{LIGOScientific:2021sio}. Therefore, it makes sense to consider theories arising as small modifications of GR and effective field theory (EFT) provides an organising principle. In EFT, one adds all possible terms allowed by symmetry to the leading order GR Lagrangian. These terms are organised in a derivative expansion and appear multiplied by dimensionful coupling constants that encode the effects of the underlying (unknown) microscopic theory. In pure gravity, after considering field redefinitions, the leading correction to GR starts at six or eight derivatives \cite{Endlich:2017tqa}.\footnote{A recent argument (that relies on certain assumptions on the UV completion of gravity) suggests that, to preserve causality in the six-derivative theories, an infinite tower of higher spin particles is required \cite{Camanho:2014apa,Serra:2022pzl}; such particles have not been observed, which would indicate that the couplings of these theories are further suppressed.} Such theories of gravity have higher order equations of motion and it is not yet understood how to obtain well-posed formulations that capture the long distance physics of interest. However, there has been promising recent progress \cite{Cayuso:2017iqc,Allwright:2018rut,Cayuso:2020lca,Franchini:2022ukz}.  

Some particular classes of higher derivative theories of gravity that have received a lot of attention in recent years are the Lovelock theories \cite{Lovelock:1971yv} in the case of pure gravity, and the Horndeski theories \cite{Horndeski:1974wa} in the case of scalar-tensor theories.\footnote{We note that Lovelock theories are only non-trivial in spacetime dimensions higher than four.}  Both Lovelock and Horndeski theories have second order equations of motion, and hence there was hope that suitable well-posed formulations could be found. In some remarkable papers, Kov\'acs and Reall have shown that these theories are indeed well-posed in a modified version of the harmonic gauge \cite{Kovacs:2020pns,Kovacs:2020ywu}. Subsequently, work has begun to study some specific scalar-tensor theories within these classes in their highly dynamical and fully non-linear regimes \cite{East:2020hgw,East:2021bqk,East:2022rqi}.

Generalized harmonic coordinates are appealing because of the manifest wave-like structure of the equations,  but their practical implementation in numerical simulations necessitates excision. The latter, whilst conceptually straightforward, can be difficult to implement in practise. As a consequence,  many groups in the numerical relativity community have opted to use singularity avoiding coordinates such as the BSSN \cite{Nakamura:1987zz,Shibata:1995we,Baumgarte:1998te}, Z4C \cite{Bona:2003fj,Bernuzzi:2009ex} or CCZ4 \cite{Alic:2011gg} formulations in the puncture gauge \cite{Campanelli:2005dd,Baker:2005vv}, which do not require the excision of the interior of black holes from the computational domain. This strongly motivates the extension of the results of \cite{Kovacs:2020pns,Kovacs:2020ywu} to singularity avoiding coordinates, to allow such groups to generate waveforms in these models.

In this Letter we modify the CCZ4 formulation of the Einstein equations together with the $1+\log$ slicing \cite{Bona:1994dr} and Gamma-driver \cite{Alcubierre:2002kk} gauge conditions and show that at least certain classes of these higher derivative scalar-tensor theories of gravity can be well-posed in singularity avoiding coordinates. Following the EFT philosophy, we consider the most general parity-invariant scalar-tensor theory of gravity up to four derivatives (4$\partial$ST) \cite{Weinberg:2008hq}:
\begin{equation}
\label{eq:action}
    \begin{aligned}
    I = \frac{1}{16\pi}\int d^4x\sqrt{-g}&\big[-V(\phi)+R+X\\
    &+g_2(\phi)X^2+\lambda(\phi)\mathcal{L}_\text{GB}\big]\,,
    \end{aligned}
\end{equation}
where $X\equiv-\frac{1}{2}(\nabla_\mu\phi)(\nabla^\mu\phi)$, $V(\phi)$ is the scalar potential, $g_2(\phi)$ and $\lambda(\phi)$ are smooth functions of the scalar field $\phi$ (but not of its derivatives), and 
\begin{equation}
    {\mathcal L}_{GB}=R^2-4R_{\mu\nu}R^{\mu\nu}+R_{\mu\nu\rho\sigma}R^{\mu\nu\rho\sigma}\,,
\end{equation}
is the Gauss-Bonnet Lagrangian.
Since well-posedness is a necessary but not sufficient condition for a stable numerical evolution, we demonstrate that our formulation works in practise by evolving a selection of binary black hole mergers.

We follow the conventions in Wald's book \cite{Wald:1984rg}. Greek letters $\mu,\nu\,\ldots$ denote spacetime indices and they run from 0 to 3; Latin letters $i,j,\ldots$ denote indices on the spatial hypersurfaces and they run from 1 to 3. We set $G=c=1$.

\section{Modified CCZ4 formulation}
\label{sec:mCCZ4}
The equations of motion (eoms) derived from \eqref{eq:action} in the modified harmonic gauge introduced by \cite{Kovacs:2020pns,Kovacs:2020ywu}, and supplemented by constraint damping terms, are given by:
\begin{align} 
&\textstyle R^{\mu\nu}-\frac{1}{2}\, R \,g^{\mu\nu}-\hat{P}_{\alpha}^{~\beta\mu\nu}\nabla_{\beta}C^{\alpha}\,
\label{eq:eom_metric}\\
&+\kappa_1\textstyle\big[n^{(\mu}C^{\nu)}+\frac{1}{2}\kappa_2\,n^{\alpha}C_{\alpha}\,g^{\mu\nu}\big]
= T^{\phi\,\mu\nu} - 4{\mathcal H}^{\mu\nu} \,,\nonumber\\
&\Box \phi[1+2g_2(\phi)X] -V'(\phi)-3X^2g_2'(\phi) \, \nonumber\\&- 2g_2(\phi)(\nabla^{\mu}\phi)(\nabla^{\nu}\phi)\nabla_{\mu}\nabla_{\nu}\phi = -\lambda'(\phi){\mathcal L}_{GB} \,, \label{eq:eom_sf}
\end{align}
where $C^{\mu}=H^{\mu}+\tilde{g}^{\rho\sigma}\Gamma_{\rho\sigma}^{\mu}$ are the constraints, and $H^{\mu}$ are the source functions that parametrize the underlying coordinate freedom of the theory; $\hat{P}_{\alpha}^{~\beta\mu\nu}=\delta_{\alpha}^{(\mu}\hat{g}^{\nu)\beta}-\frac{1}{2}\delta_{\alpha}^{\beta}\hat{g}^{\mu\nu}$ and the two additional auxiliary (inverse) metrics $\tilde g^{\mu\nu}$ and $\hat g^{\mu\nu}$ can be chosen as
\begin{align}
    \tilde{g}^{\mu\nu}=g^{\mu\nu}-a(x)n^{\mu}n^{\nu}, \ \ \hat{g}^{\mu\nu}=g^{\mu\nu}-b(x)n^{\mu}n^{\nu},
    \label{eq:aux_metrics}
\end{align}
where $n^{\mu}$ is the unit (with respect to the spacetime metric $g_{\mu\nu}$) normal to surfaces of constant $x^0$. Assuming that i) the initial data surface is spacelike for all three metrics, ii) the null cones of the three metrics do not intersect and, iii) the physical metric $g^{\mu\nu}$ has the innermost null cone, requires that $0<a(x)<b(x)$ or $0<b(x)<a(x)$. If the ordering of the null cones of $g^{\mu\nu}$ and $\tilde g^{\mu\nu}$ is interchanged, then one has the alternative condition $-1<a(x)<0<b(x)$ \cite{Kovacs:2020pns,Kovacs:2020ywu}.   The damping coefficients $\kappa_1$ and $\kappa_2$ in \eqref{eq:eom_metric} must satisfy the bounds $\kappa_1>0$ and $\kappa_2>-\frac{2}{2+b(x)}$ \cite{long_paper}. 

The terms in the right hand side (r.h.s.) of \eqref{eq:eom_metric} are given by
\begin{align}
    &T^{\phi}_{\mu\nu}=\textstyle\frac{1}{2}\big\{(\nabla_{\mu}\phi)(\nabla_{\nu}\phi)(1+2g_2(\phi)X)\, \nonumber \\&+g_{\mu\nu}\left[g_2(\phi)X^2+X-V(\phi)\right]\big\}\,, \\
    &{\mathcal H}_{\mu\nu}= \textstyle 2R^{\rho}_ {~(\mu}{\mathcal C}_{\nu)\rho}-{\mathcal C}(R_{\mu\nu}-\frac{1}{2}R\,g_{\mu\nu})-\frac{1}{2}R\,{\mathcal C}_{\mu\nu}\,\nonumber\\&+{\mathcal C}^{\alpha\beta}\left(R_{\mu\alpha\nu\beta}-g_{\mu\nu}R_{\alpha\beta}\right)\, ,
\end{align}
with 
\begin{align}
    {\mathcal C}_{\mu\nu}\equiv\lambda'(\phi)\nabla_{\mu}\nabla_{\nu}\phi+\lambda''(\phi)(\nabla_{\mu}\phi)(\nabla_{\nu}\phi)\,,
\end{align}
and ${\mathcal C}\equiv g^{\mu\nu}\mathcal{C}_{\mu\nu}$.

In (S10) of the Supplemental Material \cite{supplemental} we provide the usual 3+1 conformal decomposition of the eoms \eqref{eq:eom_metric}--\eqref{eq:eom_sf}, which were first written down in \cite{Witek:2020uzz,Julie:2020vov}. In order to evolve the system, we need to prescribe evolution equations for the gauge variables, namely the lapse $\alpha$ and shift vector $\beta^i$.   
In this Letter we propose the following generalizations of the $1+\log$ slicing condition for the lapse and Gamma-driver for the shift \cite{long_paper}:
\begin{equation}
\label{eq:mgauge}
\begin{aligned}
\partial_t\alpha &= \textstyle\beta^i\partial_i\alpha-\frac{2\alpha}{1+a(x)}(K-2\Theta)\,, \\
\partial_t\beta^i &= \textstyle\beta^j\partial_j\beta^i +\frac{3}{4} \frac{\hat{\Gamma}^i}{1+a(x)}-\frac{a(x)}{1+a(x)}\,\alpha\, D^i\alpha\,.
\end{aligned} 
\end{equation}
In this work we show that in the weakly coupled regime,  \eqref{eq:mgauge} together with (S10) form a strongly hyperbolic system of partial differential equations. The outline of the proof is provided in Section \ref{sec:well-posedness}, and the details can be found in \cite{long_paper}. In Section \ref{sec:bhbinaries} we show the robustness of our formulation by simulating black hole binaries in the theory \eqref{eq:action} for the first time.

\section{Well-posedness}
\label{sec:well-posedness}
In this section we study the well-posedness of (S10)--\eqref{eq:mgauge} in the weakly coupled regime, that is, when the higher derivative terms in the eoms are much smaller than the two-derivative ones. 

We first analyze the hyperbolicity of the two-derivative Einstein-scalar field theory in the modified CCZ4 (mCCZ4) formulation, i.e., (S10)--\eqref{eq:mgauge} with $\lambda(\phi)=g_2(\phi)=0$. We calculate the principal symbol by linearizing the equations around an arbitrary background, keeping the highest derivative terms and replacing $\partial_{\mu}\to i\xi_{\mu}\equiv i(\xi_0,\xi_i)$ with $\gamma^{ij}\xi_i\xi_j=1$. We get
\begin{equation}
    i\xi_0U = \mathbb{M}_0(\xi_k)U,
    \label{eq:ppl_part}
\end{equation}
where $U$ is the vector of all linearized variables $U=\{\hat{\tilde{\gamma}}_{ij},\hat\chi,\hat{\tilde{A}}_{ij},\hat K,\hat\Theta,\hat{\hat\Gamma}^i,\hat\alpha,\hat\beta^i,\hat\phi, \hat K_\phi\}$,\footnote{Note that taking into account the constraints $\det(\tilde{\gamma}_{ij})=1$ and ${\text Tr}(\tilde{A}_{ij})=0$, there are $22$ independent variables.} and $\mathbb{M}_0(\xi_k)$ can be found in Appendix B of the Supplemental Material \cite{supplemental}.

The system is strongly hyperbolic if $\mathbb{M}_0$ has real eigenvalues and a complete set of eigenvectors that depend smoothly on the wavevectors  $\xi_k$ for any $\xi_k$. The explicit expressions for the eigenvalues are given in \eqref{eq:eigen_phys1},\eqref{eq:eigen_phys2}, \eqref{eq:eigen_gauge_violating} and \eqref{eq:eigen_pure_gauge} setting $\lambda^\text{GB}=g_2=0$;  the eigenvectors as well as the analysis of the propagation of the constraints can be found in \cite{long_paper}.

Just as in the standard CCZ4 formulation in puncture gauge, the different sets of eigenvalues can become degenerate for specific combinations of the lapse $\alpha$ and the conformal factor $\chi$.  The degeneracy that occurs for $\alpha=\frac{1}{2\chi}$ in our mCCZ4 formulation is the same as in standard CCZ4. Additional  degeneracies occur for $\alpha=\frac{1+b(x)}{2\chi(1+a(x))}$, $\alpha^2=\frac{1+b(x)}{\chi(1+a(x))}$ and $\alpha=\frac{2(1+b(x))}{1+a(x)}$; given the ranges of $\alpha$ and $\chi$ together with the conditions that $a(x)$ and $b(x)$ satisfy, these new degeneracies can be avoided by further imposing $b(x)>\frac{1+4a(x)}{3}$ with either $0<a(x)<b(x)$ or $-1<a(x)<0<b(x)$. However, just as the degeneracy already present in the standard CCZ4 formulation does not cause problems in practical applications, nor do the new ones. Therefore, in practice we do not need to impose this extra condition on $b(x)$ to stably evolve black hole binaries, see Section \ref{sec:bhbinaries}.

To show that \eqref{eq:action} is well-posed in our mCCZ4 formulation in the weakly coupled regime we have to specify the coupling functions $\lambda(\phi)$ and $g_2(\phi)$; for simplicity, we choose: $\lambda(\phi)=\frac{\lambda^{\text{GB}}}{4}f(\phi)$ and $g_2(\phi)=g_2$, where $\lambda^{\text{GB}}$ and $g_2$ are constants that we assume to be suitably small and of the same order.\footnote{We assume $\lambda^{\text{GB}}>0$ without loss of generality.}  Then, the weakly coupled regime of the theory corresponds to
\begin{equation}
    L\gg \sqrt{\lambda'(\phi)}\,,\sqrt{|g_2|}
\end{equation}
where $L$ is any characteristic length scale of the system associated to the spacetime curvature and the gradients of the scalar field.\footnote{In practice,  $L^{-1}=\text{max}\{|R_{\mu\nu\rho\sigma}|^\frac{1}{2}$, $|\nabla_\mu\nabla_\nu\phi|^\frac{1}{2}$, $|\nabla_\mu\phi| \}$.}

The principal part of the full theory,  (S10)--\eqref{eq:mgauge}, can be written as
\begin{align}
    \mathbb{M}=\mathbb{M}_0+\delta\mathbb{M}
    \label{eq:M_full}
\end{align}
where $\delta\mathbb{M}=\lambda^{\text{GB}}\mathbb{M}^{GB}+g_2\mathbb{M}^{X}$ are the contributions from the higher derivative terms that, in the weakly coupled regime, are small compared to $\mathbb M_0$. Therefore, to prove that the full theory is well-posed in an open neighbourhood around the Einstein-scalar-field theory, we can proceed by explicitly computing the eigenvalues and eigenvectors of \eqref{eq:M_full} perturbatively and showing that $\mathbb M$ has real eigenvalues and is diagonalizable.

Consider one of the eigenvalues\footnote{Here we suppress the subscript $0$ on $\xi_0$ to simplify the notation.} of the unperturbed principal part $\mathbb{M}_0$, namely $\xi$ with multiplicity $N^\xi$; let the associated right and left eigenvectors be $\{{\bf v^{\xi}_{\text{\tiny R},i}}\}_{i=1}^{N^{\xi}}$ and $\{{\bf v^{\xi}_{\text{\tiny L},i}}\}_{i=1}^{N^{\xi}}$ respectively. The perturbed eigenvalues $\left\{\xi+\delta\zeta^{\xi}_i\right\}_{i=1}^{N^{\xi}}$ and eigenvectors $\left\{{\bf \alpha^{\xi}_i}\cdot{\bf v_{\text{\tiny R}}^{\xi}} + \delta{\bf w^{\xi}_i} \right\}_{i=1}^{N^{\xi}}$ can be obtained by solving the eigenvalue problem \cite{hinch},
\begin{align}
{\mathcal T}^{\xi}{\bf\alpha^{\xi}_i} &= i\delta\zeta^{\xi}_i{\bf\alpha^{\xi}_i} \,, \label{eq:system_hinch1}\\ 
\left(\mathbb{M}_0-i\xi{\mathbb I} \right)\delta{\bf w^{\xi}_i} &=\left(i\delta\zeta^{\xi}_i{\mathbb I}-\delta\mathbb{M} \right)({\bf \alpha^{\xi}_i}\cdot{\bf v_{\text{\tiny R}}^{\xi}})\,, \label{eq:system_hinch2}
\end{align}
where ${\mathcal T}^{\xi}_{ij}= \frac{{\bf v_{\text{\tiny L},i}^{\xi\dagger}}\delta\mathbb{M}\,{\bf v_{\text{\tiny R},j}^{\xi}}}{{\bf v_{\text{\tiny L},i}^{\xi\dagger}}{\bf v_{\text{\tiny R},i}^{\xi}}}$. Note that \eqref{eq:system_hinch1} ensures that the r.h.s. of \eqref{eq:system_hinch2} has no components parallel to $\xi$. Therefore,  the matrix  ${\mathbb M}_0-i\xi{\mathbb I}$ on the l.h.s. of \eqref{eq:system_hinch2} is invertible \cite{hinch}.

To prove well-posedness we need to verify that the matrices $\left\{{\mathcal T}^{\xi}\right\}_{\xi\in\text{Spec}({\mathbb M}_0)}$ are diagonalizable and that the perturbed eigenvectors depend smoothly on $\xi_k$. In \cite{long_paper} we prove that this holds. Shifting $\xi_0-\beta^k\xi_k\to \xi_0$, the eigenvalues of the perturbed system can be classified as follows:

\vspace{0.15cm}
\noindent
\textit{Physical eigenvalues}: Letting $\epsilon=\pm1$, the $6$ eigenvalues from this sector are split into two corresponding to the purely gravitational sector,\footnote{The corresponding eigenvectors are null with respect to the effective metric $C^{\mu\nu}=g^{\mu\nu}-4{\mathcal C}^{\mu\nu}$ as described in \cite{Reall:2021voz}.  } 
 \begin{align}
 \label{eq:eigen_phys1}
 \xi_0=&~\alpha\left(\epsilon+2\eta_{\epsilon}\right)\,,
 \end{align} 
 and four corresponding to the mixed gravitational-scalar field polarizations, 
 \begin{align}
 \label{eq:eigen_phys2}
 \xi_0=&~\alpha\left(\epsilon+\eta_{\epsilon}+\sigma_{\epsilon}\right.\\
 &\textstyle{\left.\pm\sqrt{\left(\eta_{\epsilon}-\sigma_{\epsilon}\right)^2+\psi_{12}^2+\left(\frac{\psi_{11}-\psi_{22}}{2}\right)^2}\right)},\nonumber
 \end{align}
where
\begin{align}
     \eta_{\epsilon}=&\textstyle\left[2\xi_i\gamma^i_{\mu}n_{\nu}-\epsilon\left(n_{\mu}n_{\nu} + \xi_i\xi_j\gamma^i_{\mu}\gamma^j_{\nu} \right)\right]{\mathcal C}^{\mu\nu}\,,\\
     \sigma_{\epsilon}=&~\frac{g_2}{2}\left[\xi_i(D^i\phi) K_{\phi}+\epsilon\left(K_{\phi}^2-\xi_i\xi_j(D^i\phi)(D^j\phi) \right) \right] \,,\\
     \psi_{AB}=&~\lambda^{\text{GB}}e_A^ie_B^j\Big[\textstyle{{\mathcal L}_nK_{ij}+\frac{1}{\alpha}D_iD_j\alpha} \\
     &\textstyle{+R_{ij}+KK_{ij}-K_i^{~k}K_{jk}
     +2\xi_k\big(D^kK_{ij}-D_{(i}K_{j)}^{~k} \big)}\Big]\nonumber
\end{align} 
and $\{e_A^i\}$, $A=1,2$ together with $\xi_i$ form an orthonormal triad.

\vspace{0.15cm}
\noindent
\textit{``Gauge-condition violating'' eigenvalues}: 
    \begin{equation}
    \label{eq:eigen_gauge_violating}
    \begin{aligned}
     \xi_0&=\pm\textstyle\sqrt{\frac{2\alpha}{1+a(x)}}\,,\\
     \xi_0&=\pm\textstyle\frac{1}{\sqrt{\chi(1+a(x))}}\,,\\
     \xi_0&=\pm\textstyle\frac{\sqrt{3}}{2\sqrt{\chi(1+a(x))}}\,,
    \end{aligned} 
    \end{equation}
where the last pair of eigenvalues have multiplicity $2$.

\vspace{0.15cm}\noindent
\textit{``Pure-gauge'' eigenvalues}: 
    \begin{equation}
    \label{eq:eigen_pure_gauge}
    \begin{aligned}
        \xi_0&=\pm\textstyle\frac{\alpha}{\sqrt{1+b(x)}}\,,
    \end{aligned}
    \end{equation}
where both eigenvalues have multiplicity $4$.

Clearly the eigenvalues \eqref{eq:eigen_phys1}--\eqref{eq:eigen_pure_gauge} are real (recall that in all cases $a(x)>-1$ and $b(x)>0$), they smoothly depend on $\xi_k$, and hence $\mathbb M$ is diagonalizable.  Therefore, we have shown that, in the weakly coupled regime, the system is strongly hyperbolic. 

\section{Black hole binary mergers in 4$\partial$ST}
\label{sec:bhbinaries}
We demonstrate the robustness of our proposed mCCZ4 formulation by simulating equal mass non-spinning black hole binary mergers in the theory \eqref{eq:action}. In all our simulations we have used $a(x)=0.2$ and $b(x)=0.4$, and $f(\phi)=\phi$, which corresponds to further imposing that the eoms are invariant under shifts in $\phi$. In this theory stationary black hole solutions possess a non-trivial scalar field configuration -- so called scalar hair.\footnote{References \cite{LISA:2022kgy} and \cite{Carson:2020cqb} use the same convention for the coupling as in our paper, while there is a factor of $4\sqrt{2}$ difference with respect to references \cite{Krishnendu:2021fga} and \cite{Carson:2020ter}. }

We have implemented our code as an extension to \texttt{GRChombo} \cite{Clough:2015sqa,Andrade:2021rbd}. We follow the method in \cite{Figueras:2020dzx,Figueras:2021abd} of smoothly switching off some of the higher derivative terms in the eoms well inside the apparent horizon by replacing $\lambda^{\text{GB}}\to\frac{\lambda^{\text{GB}}}{1+e^{-100(\chi-\chi_0)}}$ with $\chi_0=0.15$.\footnote{In our coordinates and for non-spinning black holes, the apparent horizon is accurately tracked by the $\chi\simeq 0.3$ contour.}
We superpose the initial perturbative solution for two GR boosted
black holes in \cite{Baumgarte:2010ndz} as initial data (and hence with vanishing scalar field) with equal masses $m_{(1)}=m_{(2)}=0.49M$, separation $11M$ and initial velocities $v_{(i)}=(0,\pm 0.09,0)$. These initial conditions have been tuned to have roughly circular initial orbits in GR such that the two black holes merge in approximately seven orbits.
We use Bowen-York initial data for the two boosted black holes \cite{Bowen:1980yu} with the Hamiltonian constraint solved approximately using a perturbative expansion in the linear momentum $P^i$ as in \cite{Baumgarte:2010ndz,Gleiser:1999hw}. No further constraint violation arises in the EFT since the additional non-GR terms in the constraints vanish for zero initial scalar field \cite{East:2020hgw,Ripley:2022cdh}. 
 As the scalar hair grows and the punctures form, constraint violations become visible inside the apparent horizon (both due to the puncture and the switching off of the coupling terms described above). Outside the apparent horizon and in the weakly coupled regime they stay small provided that the damping terms in the eoms are chosen appropriately.
For the simulations shown here, the values of the constraint damping coefficients have been set to $\kappa_1=0.35$ and $\kappa_2=-0.1$.

\begin{figure}[t]
\centering
\includegraphics[width=9cm]{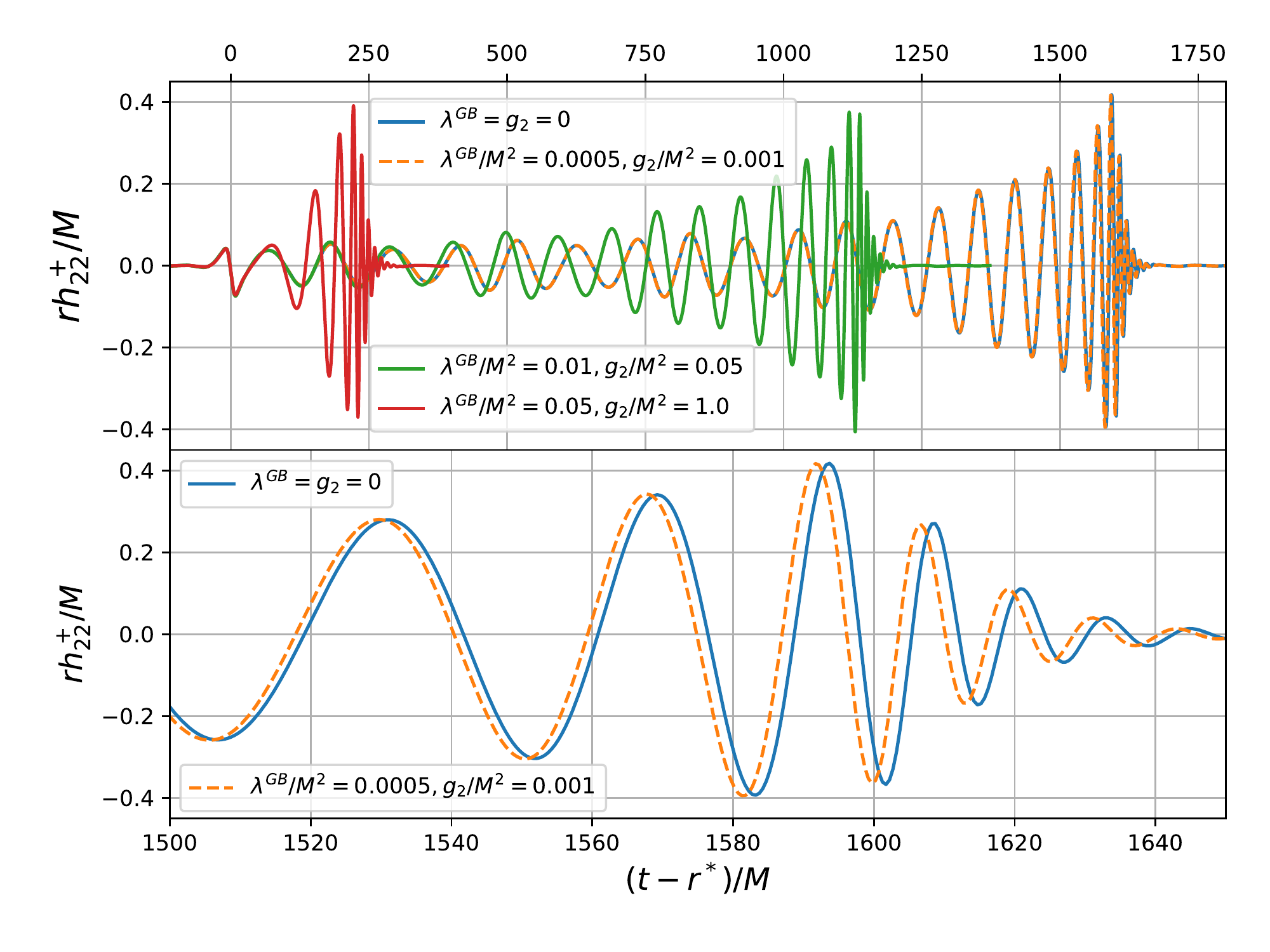}
\caption{\textit{Top}: Comparison of the (2,2) mode of the gravitational wave strain between GR (blue) and $4\partial$ST in retarded time, $u=t-r^*$, where $r^*$ is the tortoise coordinate, for different values of the couplings, namely small (orange dashed), medium (green) and large (red). \textit{Bottom}: Zoom in of the merger region for the small coupling case.}
\label{fig:all_waves}
\end{figure}

\begin{figure}[t]
\centering
\includegraphics[width=9cm]{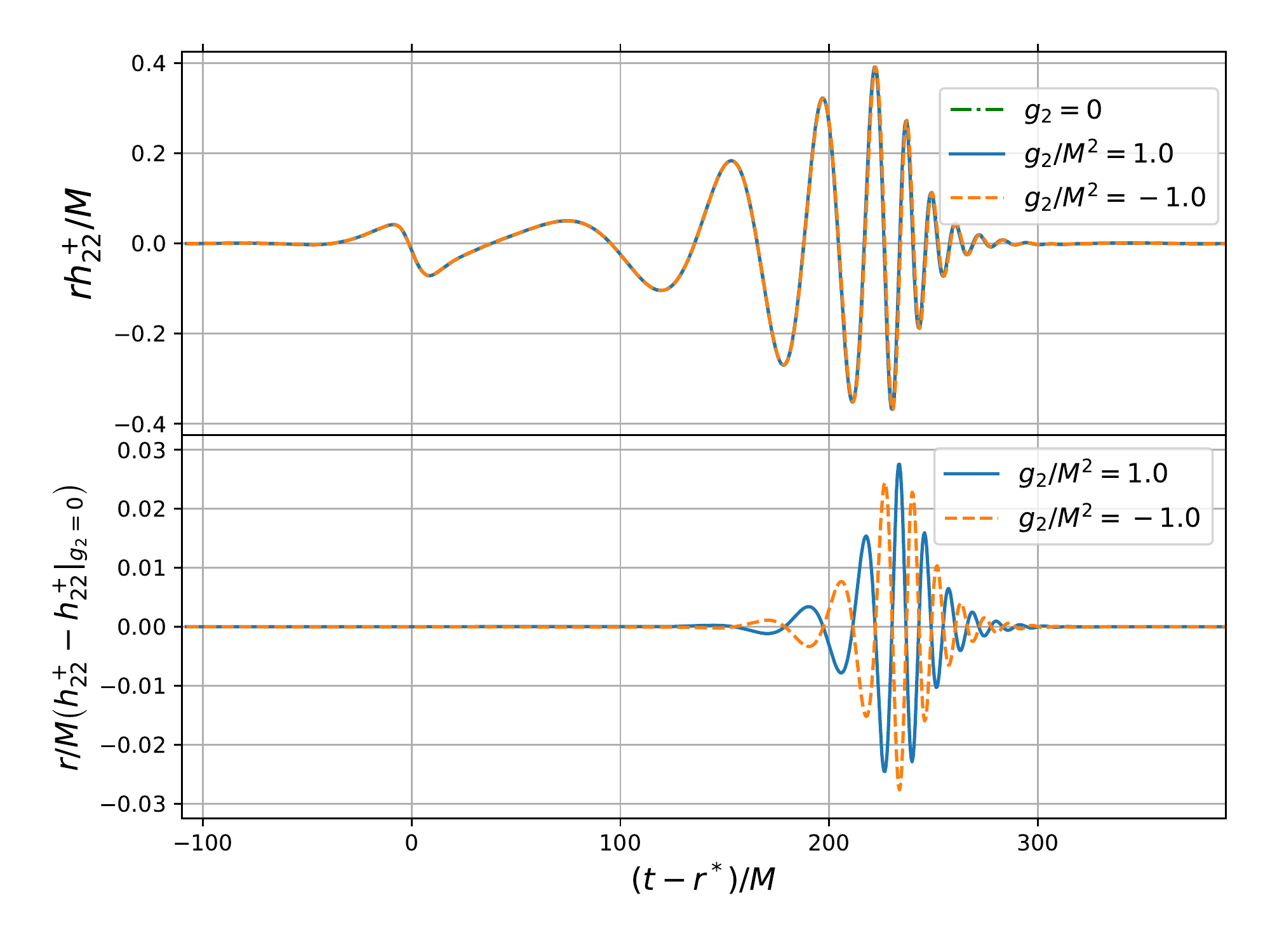}
\caption{\textit{Top}: Waveforms for fixed $\lambda^{\text{GB}}/M^2=0.05$, with different values of the Horndeski coupling $g_2$. We see that changing $g_2$ has a small effect since the dynamics are mainly controlled by the Gauss-Bonnet coupling $\lambda^\text{GB}$. \textit{Bottom}: The lower plot shows the difference between the strains for the large couplings $g_2/M^2=\pm 1$ compared to the $g_2=0$ case.}
\label{fig:large_coupling}
\end{figure}

For each simulation that we have run, we have extracted the gravitational waves at $r=100M$ and computed the $(\ell,m)=(2,2)$ mode of the plus polarization of the strain, $h_{22}^+$, for comparison. The results are shown in Figs. \ref{fig:all_waves}, \ref{fig:large_coupling}. Since in the theory that we consider the Gauss-Bonnet term sources the scalar field, it is the associated coupling constant $\lambda^\text{GB}$ that plays the most prominent role and effectively controls the regime of validity of the EFT. For very small values of the couplings, the waveforms in $4\partial$ST tend to GR, as expected since the scalar field is effectively perturbative.\footnote{We emphasize that we still treat the system non-perturbatively, which allows us to avoid potential issues with secular effects.} The differences only become noticeable in the merger phase (see the bottom panel in Fig. \ref{fig:all_waves}) and they appear as a phase shift in the waveform. This is expected since the spacetime curvature outside black holes is largest during this phase. 

The situation is drastically different for large values of the couplings.  For $\lambda^\text{GB}/M^2=0.05$, the binary merges in only 3 orbits as opposed to the 7 orbits that the black holes describe in GR with the same initial conditions. In the $4\partial$ST theory with large couplings the system can radiate strongly in scalar waves and hence shed energy and angular momentum more efficiently than in GR, so the larger the $\lambda^\text{GB}/M^2$ coupling, the sooner the binary merges \cite{Yagi:2011xp,Shiralilou:2020gah,Shiralilou:2021mfl}. Furthermore, the formation of the scalar cloud from an initial zero state may have an effect on the circular orbits which needs to be carefully quantified.
We were able to increase the coupling to $\lambda^\text{GB}/M^2=0.1$ without major difficulties where we find that the binary merges even quicker. It seems possible to increase this coupling even further, but each increase necessitates a tuning of the damping parameters $\kappa_1$ and $\kappa_2$ to keep the truncation errors under control, and so we leave a full exploration of the limit to future work.

In Fig. \ref{fig:large_coupling} we compare the effect of varying $g_2$ for a fixed (large) value of $\lambda^\text{GB}$. Even if the values of $g_2$ are large compared to the values used in \cite{Figueras:2021abd}, the effect is small. The reason is that the typical energy densities of the scalar field that result from the scalarization process in the 4$\partial$ST are much smaller than in \cite{Figueras:2021abd}. In the present case we have $\sqrt{|g_2|}/L \lesssim 0.1 $ while $0.2\lesssim \sqrt{\lambda^{GB}}/L\lesssim 0.5$ throughout the evolution. It is interesting to note that changing the sign of $g_2$ gives rise to a phase shift of roughly $\pi$ rather than an advancement or delay of the wave as in \cite{Figueras:2021abd}.

\section{Discussion}
\label{sec:discussion}
In this article we have proposed a modified version of the CCZ4 formulation of the Einstein equations based on \cite{Kovacs:2020pns,Kovacs:2020ywu}, together with a modification of the puncture gauge extensively used in numerical relativity to simulate black hole binary mergers. With these modifications we have proved the well-posedness of the most general scalar-tensor theory of gravity up to four derivatives in the weakly coupled regime in singularity avoiding coordinates. In \cite{long_paper} we also prove well-posedness of the Lovelock theory of gravity using our formulation.  It seems plausible that one can also extend the well-posedness results in singularity avoiding coordinates to the general Horndeski theory - see the argument in \cite{Kovacs:2020pns,Kovacs:2020ywu}, which avoids the explicit computation of the eigenvalues and eigenvectors of the theory. 

We have demonstrated the robustness of our formulation in practise by simulating black hole binary mergers in the $4\partial\text{ST}$ theory \eqref{eq:action} for different values of the coupling constants, with values up to $\lambda^{GB}/M^2=0.1$ remaining stable for the evolution, and compared the waveforms to those of GR. Treating the theory fully non-linearly allows us to avoid the secular effects of order reduced formulations while capturing the non-perturbative physics. The dynamics of our model is essentially controlled by the Gauss-Bonnet coupling $\lambda^{GB}$, since it is the Gauss-Bonnet term that gives rise to the scalar hair of the black holes.  For large values of $\lambda^{GB}$, i.e., $\lambda^{GB}\simeq 0.05/M^2$, the deviations from GR are large, as one would expect since the $4\partial\text{ST}$ theory has a scalar degree of freedom. However, even for these large values of the couplings, the theory is weakly coupled. It would be interesting to check the limits in which one can probe the strong coupling regime of the theory with our formulation.   

The focus of this Letter is on the success of the methods rather than the phenomenology of the models that we have considered. Our work, building on previous studies \cite{Kovacs:2020pns,Kovacs:2020ywu,East:2020hgw}, opens up the possibility for detailed study of phenomenologically interesting scalar-tensor theories of gravity with two-derivative equations of motion to the many existing numerical relativity codes that rely on puncture gauges. These codes, many of which are crucial for current LIGO-Virgo-KAGRA inference, now have the means to obtain high quality waveform templates in such theories. These template banks can help contrast the predictions of specific ``beyond GR'' models with parameterised approaches as well as present and future gravitational wave observations.

\subsection*{Acknowledgements}
We would like to thank Aron Kov\'acs for numerous discussions about well-posedness. We thank Harvey Reall for his helpful comments on the nature of the various polarizations. We also want to thank Tiago Fran\c{c}a for his support in some of the technical aspects of the paper and Ulrich Sperhake and Miren Radia for help with the initial GR binary parameters. 
We want to thank the entire \texttt{GRChombo}\footnote{\texttt{www.grchombo.org}} collaboration for their support and code development work. PF would like to thank the Enrico Fermi Institute and the Department of Physics of the University of Chicago for hospitality during the final stages of this work. This work was first presented at the workshop ``Frontiers in Numerical Relativity 2022'' at the University of Jena; we would like to thank the organizers for inviting us and the participants for stimulating discussions. PF is supported by a Royal Society University Research Fellowship  No. URF\textbackslash R\textbackslash 201026, and No. RF\textbackslash ERE\textbackslash 210291. KC is supported by an STFC Ernest Rutherford fellowship, project reference ST/V003240/1. LAS is supported by a QMUL Ph.D. scholarship. The simulations presented used PRACE resources under Grant No. 2020235545, PRACE DECI-17 resources under Grant No. 17DECI0017, the CSD3 cluster in Cambridge under Projects No. DP128. The Cambridge Service for Data Driven Discovery (CSD3), partially operated by the University of Cambridge Research Computing on behalf of the STFC DiRAC HPC Facility. The DiRAC component of CSD3 is funded by BEIS capital via STFC capital Grants No. ST/P002307/1 and No. ST/ R002452/1 and STFC operations Grant No. ST/R00689X/1. DiRAC is part of the National e-Infrastructure.\footnote{\texttt{www.dirac.ac.uk}} The authors gratefully acknowledge the Gauss Centre for Supercomputing e.V.\footnote{\texttt{www.gauss-centre.eu}} for providing computing time on the GCS Supercomputer SuperMUC-NG at Leibniz Supercomputing Centre.\footnote{\texttt{www.lrz.de}} Calculations were performed using the Sulis Tier 2 HPC platform hosted by the Scientific Computing Research Technology Platform at the University of Warwick. Sulis is funded by EPSRC Grant EP/T022108/1 and the HPC Midlands+ consortium. This research also
utilised Queen Mary’s Apocrita HPC facility, supported
by QMUL Research-IT \cite{apocrita}. For some computations we have also used the Young Tier 2 HPC cluster at UCL; we are grateful to the UK Materials
and Molecular Modelling Hub for computational resources, which is partially funded by
EPSRC (EP/P020194/1 and EP/T022213/1). For the purpose of Open Access, the author has applied a CC BY public copyright licence to any Author Accepted Manuscript version arising from this submission.

\phantom{\cite{ccz4,ccz42,Radia:2021smk}}

\appendix
\section{Equations of motion}
\label{app:full_eoms}
In this section we write down the equations of motion of the theory, eqs. \eqref{eq:eom_metric}--\eqref{eq:eom_sf}, in the 3+1 form as we have implemented in our code.  We consider the usual $3+1$ decomposition of the spacetime metric, 
\begin{equation}
    ds^2=-\alpha^2\,dt^2+\gamma_{ij}(dx^i+\beta^i\,dt)(dx^j+\beta^j\,dt)\,,
\end{equation}
where $\alpha$ and $\beta^i$ are the lapse function and shift vector respectively, and $\gamma_{ij}$ is the induced metric on the $t\equiv x^0=\text{const}.$ hypersurfaces. In these coordinates, the unit timelike vector normal to these hypersurfaces is given by $n^{\mu}=\frac{1}{\alpha}(\delta_t^{\mu}-\beta^i\delta_i^{\mu})$ and the extrinsic curvature is given by
\begin{equation}
    K_{\mu\nu} = -\tfrac{1}{2}\mathcal{L}_n \gamma_{\mu\nu}\,,
\end{equation}
where $\mathcal{L}_n$ denotes the Lie derivative along $n^\mu$.

The evolution variables are decomposed as,
\begin{align}
\chi&=\det(\gamma_{ij})^{-\frac{1}{3}}, \\
\tilde{\gamma}_{ij}&=\chi\,\gamma_{ij}, \\ \tilde{A}_{ij}&=\chi\left(K_{ij}-\frac{1}{3}\,\gamma_{ij}K\right)\,,  \\
\hat{\Gamma}^i&=\tilde{\Gamma}^i+2\tilde{\gamma}^{ij}Z_j\,  ,
\end{align}
where $\tilde{\Gamma}^i\equiv\tilde\gamma^{kl}\tilde \Gamma^i_{kl}$, and $\tilde \Gamma^i_{kl}$ are the Christoffel symbols associated to the conformal spatial metric $\tilde{\gamma}_{ij}$.\footnote{Note that the conformal spatial metric $\tilde \gamma_{ij}$ is unrelated to the auxiliary spacetime metric $\tilde g_{\mu\nu}$ defined in \eqref{eq:aux_metrics}.} Note that the spatial indices are raised and lowered with the physical spatial metric $\gamma_{ij}$. In addition, we have the components of the Z4 vector \cite{ccz4,ccz42},\footnote{In our conventions, the Z4 vector $Z^\mu$ is related to the vector of constraints $C^\mu$ as $Z^\mu = \frac{1}{2}C^\mu$.} 
\begin{align}
    \Theta&\equiv Z^0=\tfrac{1}{2}\,C^{\perp}\,, \\
    Z^i&=-\tfrac{1}{2}\,C^i\,,
\end{align}
where
\begin{subequations}\label{constraints}
\begin{eqnarray}
&C^{\perp}=H^{\perp}+K+\frac{1}{\alpha}(1+a(x)){\mathcal L}_n\alpha, \\
&\hspace{-0.5cm} C_i=H_i + \Gamma_i - \frac{1+a(x)}{\alpha}\left[D_i\alpha +\frac{\gamma_{ij}}{\alpha}(\partial_t-\beta^k\partial_k)\beta^j\right]\,,
\end{eqnarray}
\end{subequations}
where $C^{\perp}\equiv n_\mu C^\mu$,  $\Gamma^{k}_{ij}$ are the Christoffel symbols of the spatial metric and $\Gamma_i\equiv \gamma_{ij}\gamma^{kl}\Gamma^{j}_{kl}$.

Having defined all the variables, their evolution equations are given by:
\begin{subequations}
\begin{eqnarray}\label{eq:mccz4_eqs}
\partial_\perp\tilde{\gamma}_{ij} &=& -2\alpha\tilde{A}_{ij}+2\tilde\gamma_{k(i}\partial_{j)}\beta^k-\tfrac{2}{3}\tilde{\gamma}_{ij}(\partial_k\beta^k), \\
\partial_\perp\chi &=& \tfrac{2}{3}\chi\left(\alpha\,K - \partial_k\beta^k\right), \\
\partial_\perp\Theta &=&\tfrac{1}{1+b(x)}\bigg\{ \tfrac{1}{2}\alpha\Big[R+2(1+b(x))D_iZ^i-\tilde{A}_{ij}\tilde{A}^{ij}\nonumber\\
&&\hspace{2cm}+\tfrac{2}{3}\,K^2
 -2(1+b(x))\,K\,\Theta \Big] \nonumber\\
&&\hspace{1.5cm}-\left(\tfrac{2+b(x)}{2}\right)Z^i\partial_i\alpha\nonumber\\
&&\hspace{1.5cm}-\alpha\kappa_1(2+\kappa_2)\Theta\nonumber\\
&&\hspace{1.5cm}-\tfrac{1}{4}\alpha\left[K_{\phi}^2\big(1+\tfrac{3}{2}g_2\left(K_{\phi}^2-(\partial\phi)^2\right) \big)\right.\nonumber\\
&&\hspace{2cm}\left.+(\partial\phi)^2\big(1+\tfrac{1}{2}g_2\left(K_{\phi}^2-(\partial\phi)^2\right) \big)\right]\nonumber\\
&&\hspace{1.5cm}-\alpha\,\lambda^{\text{GB}}\rho^{GB}\bigg\}\,,\\
\partial_\perp\hat{\Gamma}^i &=&-2\,\tilde{A}^{ij}\partial_j\alpha\nonumber\\
&&+\tfrac{2\alpha}{1+b(x)}\bigg[(1+b(x))\tilde{\Gamma}^i_{jk}\tilde{A}^{jk}-\frac{2}{3}\tilde{\gamma}^{ij}\partial_jK\nonumber\\
&&\hspace{1.5cm}-\left(\tfrac{3+5b(x)}{2\chi}\right)\tilde{A}^{ij}\partial_j\chi\bigg]\nonumber\\
&&+\tilde{\gamma}^{kl}\partial_k\partial_l\beta^i+ \tfrac{1}{3}\tilde{\gamma}^{ik}\partial_k\partial_l\beta^l\nonumber\\
&&+\tfrac{2}{3}\hat \Gamma^i(\partial_k\beta^k)-\hat\Gamma^j\partial_j\beta^i\nonumber\\
&&+\tfrac{2}{1+b(x)}\tilde{\gamma}^{ij}\bigg(\alpha\,\partial_j\Theta-(1+b(x))\,\Theta\,\partial_j\alpha\nonumber\\
&&\hspace{2cm}-\left(\tfrac{2+b(x)}{3}\right)\alpha\,K\,Z_j\bigg)\nonumber\\
&&-2\,\alpha\,\kappa_1\tilde{\gamma}^{ij}Z_j\nonumber\\
&&+ \left(\tfrac{2\, b(x)}{1+b(x)}\right)\alpha(\tilde{A}^{ij}Z_j-D_j\tilde{A}^{ij})\nonumber\\
&&-\left(\tfrac{1}{1+b(x)}\right)\alpha\,K_{\phi}\tilde{\gamma}^{ij}\partial_j\phi\big(1+g_2(K_{\phi}^2-(\partial\phi)^2)\big)\nonumber\\
&&-\left(\tfrac{2\,\,\lambda^{\text{GB}}}{1+b(x)}\right)\alpha\,\tilde{\gamma}^{ij}J^{GB}_j\,,\\
\partial_\perp\phi &=& -\alpha\,K_\phi \,,
\end{eqnarray}
\end{subequations}
where $\partial_\perp=\partial_t-\beta^i\partial_i$,  $(\partial\phi)^2\equiv \gamma^{ij}(\partial_i\phi)(\partial_j\phi)$ and
\begin{subequations}\label{edgbcomp}
\begin{eqnarray}
\rho^{GB}&=&\frac{\Omega M}{2} - M_{kl}\Omega^{kl}, \, \\
J^{GB}_i&=&\frac{\Omega_iM}{2}-M_{ij}\Omega^j - 2\left(\Omega^j_ {~[i}N_{j]}-\Omega^{jk}D_{[i}K_{j]k}\right),\, \hspace{0.8cm}
\end{eqnarray}
\end{subequations}
with
\begin{subequations}
\begin{eqnarray}
M_{ij}&=&R_{ij}+\tfrac{1}{\chi}\left(\tfrac{2}{9}\tilde{\gamma}_{ij}K^2+\tfrac{1}{3}K\tilde{A}_{ij}-\tilde{A}_{ik}\tilde{A}_j^{~k} \right), \\
N_i&=&\tilde{D}_j\tilde{A}_i^{~j}-\tfrac{3}{2\chi}\tilde{A}_i^{~j}\partial_j\chi-\tfrac{2}{3}\partial_iK, \\
\Omega_i&=&f'\left(\partial_iK_{\phi}-(\partial_j\phi)\tilde{A}^j_{~i}-\tfrac{1}{3}K\partial_i\phi \right)\\&&+f''K_{\phi}\partial_i\phi\nonumber,\\
\Omega_{ij}&=&f'\left(D_iD_j\phi-K_{\phi}K_{ij}\right)+f''(\partial_i\phi) \partial_j\phi
\end{eqnarray}
\end{subequations}
and 
\begin{subequations}
\begin{align}
    M^{\text{TF}}_{ij} &\equiv M_{ij}-\tfrac{1}{3}\gamma_{ij}M\,,\\
    \Omega^{\text{TF}}_{ij} &\equiv \Omega_{ij}-\tfrac{1}{3}\gamma_{ij}\Omega\,,
\end{align}
\end{subequations}
where $M=\gamma^{kl}M_{kl}$ and $\Omega=\gamma^{kl}\Omega_{kl}$. 

The remaining variables satisfy the following system of coupled partial differential equations:
\begin{eqnarray}
\begin{pmatrix} X^{kl}_{ij} & Y_{ij} & 0 \\ X^{kl}_K & Y_K & 0 \\ X^{kl}_{K_{\phi}} & Y_{K_{\phi}} & I \end{pmatrix}
\begin{pmatrix} \partial_t \tilde{A}_{kl} \\ \partial_tK \\ \partial_tK_{\phi} \end{pmatrix}=
\begin{pmatrix} Z_{ij}^{\tilde{A}} \\ Z^K \\ Z^{K_{\phi}} \end{pmatrix},
\end{eqnarray}
where the elements of the matrix are defined as follows,
\begin{subequations}
\begin{eqnarray}
X_{ij}^{kl}&=& \gamma_i^k\gamma_j^l\left(1-\frac{\lambda^{\text{GB}}}{3}\Omega \right)+2\lambda^{\text{GB}}\left(\gamma_{(i}^k\Omega_{j)}^{\text{TF},l}\right.\nonumber\\&&\left. -\frac{\gamma_{ij}\Omega^{\text{TF},kl}}{3}-\lambda^{\text{GB}}f'^2M^{\text{TF}}_{ij}M^{\text{TF},kl}\right) ,\\
X_{K_{\phi}}^{kl}&=&\frac{\lambda^{\text{GB}}}{2\chi}f'M^{\text{TF},kl}, \\
Y_{ij}&=&\frac{\lambda^{\text{GB}}}{3}\chi\left(-\Omega^{\text{TF}}_{ij}+\lambda^{\text{GB}}f'^2MM^{\text{TF}}_{ij} \right) , \\
Y_K&=&1+\frac{\lambda^{\text{GB}}}{3}\left(-\Omega+\frac{\lambda^{\text{GB}}}{4}f'^2M^2 \right),\\
Y_{K_{\phi}}&=&-\frac{\lambda^{\text{GB}}}{12}f'M , \\
I&=&1+g_2(3K_{\phi}^2-(\partial\phi)^2) \,,
\end{eqnarray}
\end{subequations}
while the terms on the r.h.s. are
\begin{subequations}
\begin{eqnarray}
Z_{ij}^{\tilde{A}}&=&\chi\big[-D_iD_j\alpha+\alpha\left(R_{ij} + 2D_{(i}Z_{j)} \right.\nonumber\\&&\left.-\tfrac{1}{2}(D_i\phi)D_j\phi\big(1+g_2\left(K_{\phi}^2-(D_i\phi)^2\right) \big)\right) \big]^{\text{TF}}\nonumber\\
&&+\alpha\left[\tilde{A}_{ij}(K-2\Theta)-2\tilde{A}_{il}\tilde{A}^l_{~j}\right]\nonumber\\
&&+\beta^k\partial_k\tilde{A}_{ij}+2\,\tilde A_{k(i}\partial_{j)}\beta^k-\tfrac{2}{3}\tilde{A}_{ij}(\partial_k\beta^k)\nonumber\\
&&+\lambda^{\text{GB}}\,\chi\, \tilde{S}^{GB}_{ij}\,,\\
Z^{K}&=&-D^iD_i\alpha + \left(\tfrac{1}{1+b(x)}\right)\alpha\big[R+2D_iZ^i+K(K-2\Theta)\nonumber\\
&&\hspace{0.5cm}+b(x)\big(\tfrac{1}{4}R+2D_iZ^i+\tfrac{3}{4}\tilde{A}_{ij}\tilde{A}^{ij}+\tfrac{1}{2}K^2-2\Theta K\big)\big]\nonumber\\
&&+\beta^i\partial_iK-\left(\tfrac{3\,\kappa_1}{2(1+b(x))}\right)\big(2+\kappa_2(2+b(x))\big)\,\alpha\,\Theta \nonumber\\
&&-\left(\tfrac{1}{2(1+b(x))}\right)\,\alpha\big[(\partial\phi)^2+\tfrac{1}{4}b(x)\big((\partial\phi)^2-3\,K_\phi^2 \big) \big]\nonumber\\
&&+\alpha g_2\frac{b(x)-2}{16(1+b(x))}\left(K_{\phi}^2-(\partial\phi)^2\right)\left((\partial\phi)^2+3K_{\phi}^2 \right)\nonumber\\
&&+\lambda^{\text{GB}}\left[S_K^{GB}+\left(\tfrac{3\,b(x)}{2(1+b(x))}\right)\,\alpha\,\rho^{GB}\right]\,,\\
Z^{K_{\phi}}&=& \alpha\big(-D^iD_i\phi+K\,K_\phi\big)\big(1+g_2\left(K_{\phi}^2-(\partial\phi)^2\right) \big)\nonumber\\&&+\beta^i\partial_i K_\phi\big(1+g_2\left(3K_{\phi}^2-(\partial\phi)^2\right) \big)\nonumber\\&&- (D^i\phi) \left[D_i\alpha\big(1+g_2\left(3K_{\phi}^2-(\partial\phi)^2\right) \big)   \right.\nonumber\\
&&\left.+2\alpha g_2 K_{\phi}\left(2D_iK_{\phi}-D^j\phi\frac{\tilde{A}_{ij}}{\chi}-\frac{K}{3}D_i\phi\right)\right]\nonumber\\&&+2\alpha g_2(D^i\phi) (D^j\phi) D_iD_j\phi-\tfrac{\lambda^{\text{GB}}}{4}\,f'\,S^{GB}_{K_{\phi}}\,,
\end{eqnarray}
\end{subequations}
where
\begin{subequations}
\begin{eqnarray}
\tilde{S}_{ij}^{GB}&=&\tfrac{1}{3}\left(\Omega^{TF}_{ij}-\lambda^{\text{GB}}f'^2MM^{\text{TF}}_{ij}\right)\nonumber\\
&&\hspace{0.5cm}\times\big[-\beta^i\partial_iK
+D_iD^i\alpha-\alpha\left(\tilde{A}_{kl}\tilde{A}^{kl}+\tfrac{1}{3}\,K^2\right) \big]\nonumber\\
&&+\alpha \,M_{ij}^{TF}\left[\Omega+f''(K_{\phi}^2-(\partial\phi)^2)-\lambda^{\text{GB}}f'^2H \right]\nonumber\\
&&+\tfrac{1}{3}\,\Omega\left[D_iD_j\alpha+\tfrac{1}{\chi}\big(\alpha\,\tilde{A}_{im}\tilde{A}^m_{~j}-\hat{\theta}_{ij}\big) \right]^{\text{TF}} \nonumber\\
&&-2\,\Omega_{(i}^{\text{TF},k}\left[D_{j)}D_k\alpha+\tfrac{1}{\chi}\big(\alpha\,\tilde{A}_{j)m}\tilde{A}_k^{~m}-\hat{\theta}_{j)k}\big) \right] \nonumber\\
&&+\tfrac{2}{3}\,\Omega_{ij}^{\text{TF}}\left(D_kD^k\alpha-\alpha\,\tilde{A}_{kl}\tilde{A}^{kl} \right)-\alpha\left[N_{(i}\Omega_{j)}\right]^{\text{TF}}\nonumber\\
&&+2\,\left(\tfrac{1}{3}\,\gamma_{ij}\,\Omega^{\text{TF},kl}+\lambda^{\text{GB}}f'^2M_{ij}^{\text{TF}}M^{\text{TF},kl} \right)\nonumber\\
&&\hspace{0.5cm}\times\left[D_kD_l\alpha
+\tfrac{1}{\chi}\big(\alpha\,\tilde{A}_{km}\tilde{A}^m_{~l}-\hat{\theta}_{kl}\big) \right]\nonumber\\
&&+\alpha\left[2\left(D_kA_{ij}-D_{(i}A_{j)k}\right)\Omega^k\right.\nonumber\\
&&\hspace{0.75cm}\left.+\gamma_ {ij}\,(D^kA_{kl})\,\Omega^l-\Omega_{(i}D^kA_{j)k} \right] \,,\\
S^{GB}_K&=&\tfrac{1}{3}\left(\Omega-\tfrac{\lambda^{\text{GB}}}{4}f'^2M^2\right)\nonumber\\
&&\hspace{0.75cm}\times\left[-\beta^i\partial_iK+D_iD^i\alpha- \alpha\left(\tilde{A}_{ij}\tilde{A}^{ij}+\tfrac{1}{3}\,K^2 \right) \right]\nonumber\\
&&+\alpha\, M\left(\tfrac{1}{4}\,f''(K_{\phi}^2-(\partial\phi)^2)-\tfrac{\lambda^{\text{GB}}}{4}f'^2H+\tfrac{1}{3}\,\Omega\right)\nonumber\\
&&-\alpha\left(\Omega^iN_i+\tfrac{1}{2}\,\Omega^{\text{TF},ij}\,M^{\text{TF}}_{ij} \right)-2\alpha\rho^{GB}
\nonumber\\
&&-\tfrac{1}{2}\big(\Omega^{\text{TF},kl}-\lambda^{\text{GB}}f'^2MM^{\text{TF},kl}\big)\nonumber\\
&&\hspace{0.75cm}\times\left(D_kD_l\alpha+\tfrac{\alpha}{\chi}\tilde{A}_{km}\tilde{A}_{~l}^m -\frac{\hat{\theta}_{kl}}{\chi} \right)\,,  \\
S^{GB}_{K_{\phi}}&=&-\tfrac{4}{3}\,M\left[-\beta^i\partial_iK + D_iD^i\alpha-\alpha\left(\tilde{A}_{ij}\tilde{A}^{ij}+\tfrac{1}{3}\,K^2\right) \right] \nonumber\\
&&+8\,M^{\text{TF},kl}\left[D_kD_l\alpha+\tfrac{1}{\chi}\left(\alpha\,\tilde{A}_{kj}\tilde{A}^j_{~l}-\hat{\theta}_{kl}\right) \right]\nonumber\\
&&-4\,\alpha\, H\, ,
\end{eqnarray}
\end{subequations}
where we have used $\hat{\theta}_{kl}={\mathcal L}_{\beta}\tilde{A}_{kl}+\tfrac{2}{3}\left(\alpha K-\partial_i\beta^i \right)\tilde{A}_{kl}$ with ${\mathcal L}_{\beta}\tilde{A}_{ij}=\beta^k\partial_k \tilde A_{ij}+2\tilde A_{k(i}\partial_{j)}\beta^k$, and 
\begin{equation}
    \begin{aligned}
     H=&-\tfrac{4}{3}D_iK\left(N^i+\frac{D^iK}{3}\right) \\
    &+2\,D_iA_{jk}\left(D^iA^{jk}-D^jA^{ik} \right)\,.
    \end{aligned}
\end{equation}

\section{Principal part of the Einstein-scalar sector}
\label{app:hyp_mat}

In this section we write down the principal part of the Einstein-scalar field theory in the modified CCZ4 formulation. Defining $\check{\xi}_0=\xi_0-\beta^i\xi_i$, we have:
\begin{subequations}\label{shypccz4}
\begin{eqnarray}
i\check{\xi}_0\hat{\tilde{\gamma}}_{ij} &=& 2i\tilde{\gamma}_{k(i}\xi_{j)}\hat{\beta}^k-2\alpha\hat{\tilde{A}}_{ij}-\frac{2i}{3}\tilde{\gamma}_{ij}\xi_k\hat{\beta}^k\,, \\
i\check{\xi}_0\hat{\chi} &=&\frac{2}{3}\chi\left(\alpha\hat{K}-i\xi_k\hat{\beta}^k \right), \\
i\check{\xi}_0\hat{\phi} &=&-\alpha \hat{K}_{\phi}\,, \\
i\check{\xi}_0\hat{K} &=& \hat{\alpha} +\frac{3b(x)\alpha}{4(1+b(x))}\left(\frac{\xi^l\xi^k\hat{\tilde{\gamma}}_{kl}}{\chi}\right.\nonumber\\&&\left.-\frac{\tilde{\gamma}^{jk}\hat{\tilde{\gamma}}_{jk}}{2} \right)+ i\alpha\chi \xi_i\hat{\hat{\Gamma}}^i  -\left(\frac{2}{\chi}\hat{\chi}\right.\nonumber\\&&\left.-\frac{\tilde{\gamma}^{ij}\hat{\tilde{\gamma}}_{ij}}{2} \right) \frac{\alpha(4+b(x))}{4(1+b(x))}\,, \\
i\check{\xi}_0\hat{K}_{\phi} &=& -\alpha\hat{\phi}\,, \\
i\check{\xi}_0\hat{\Theta} &=& -\frac{\alpha}{2(1+b(x))}\left(\frac{2}{\chi}\hat{\chi}-\frac{\tilde{\gamma}^{ij}\hat{\tilde{\gamma}}_{ij}}{2} \right) + \frac{i\alpha\chi \xi_i\hat{\hat{\Gamma}}^i}{2}\nonumber\\&& +\frac{b(x)\alpha}{2(1+b(x))}\left(\frac{\xi^l\xi^k\hat{\tilde{\gamma}}_{kl}}{\chi}-\frac{\tilde{\gamma}^{jk}\hat{\tilde{\gamma}}_{jk}}{2} \right)\,, \\
i\check{\xi}_0\hat{\tilde{A}}_{ij} &=& \left(\xi_i\xi_j - \frac{1}{3}\frac{\tilde{\gamma}_{ij}}{\chi}\right)\left(\chi\hat{\alpha} - \frac{\alpha}{2}\hat{\chi}\right)\nonumber \\&&+i\alpha\chi\left(\tilde{\gamma}_{k(i}\xi_{j)}\hat{\hat{\Gamma}}^k-\frac{\tilde{\gamma}_{ij}\xi_k\hat{\hat{\Gamma}}^k}{3} \right)\nonumber\\&&+\frac{1}{2}\alpha \left(\hat{\tilde{\gamma}}_{ij}-\frac{\tilde{\gamma}_{ij}\tilde{\gamma}^{kl}\hat{\tilde{\gamma}}_{kl}}{3} \right)\,, \\
i\check{\xi}_0\hat{\hat{\Gamma}}^i &=& \frac{i\alpha}{\chi(1+b(x))}\left(-\frac{4}{3}\xi^i\hat{K}-2\chi b(x)\xi_j\hat{\tilde{A}}^{ij}\right.\nonumber\\&&\left.+ 2\xi^i\hat{\Theta} \right) - \frac{1}{\chi}\left(\hat{\beta}^i + \frac{1}{3}\xi^i\xi_l\hat{\beta}^l \right)\,,\\
i\check{\xi}_0\hat{\alpha} &=& -\frac{2\alpha}{1+a(x)} (\hat{K}-2\hat{\Theta})\,, \\
i\check{\xi}_0\hat{\beta}^i &=& \frac{3}{4(1+a(x))}\hat{\hat{\Gamma}}^i-\frac{ia(x)}{1+a(x)}\alpha \xi^i\hat{\alpha}\,, 
\end{eqnarray}
\end{subequations}
Comparing with \eqref{eq:ppl_part}, one can identify the non-vanishing components of $\mathbb{M}_0(\xi_k)$ from the r.h.s. of \eqref{shypccz4}.

\section{Convergence}

\begin{figure}[t]
\centering
\includegraphics[width=9cm]{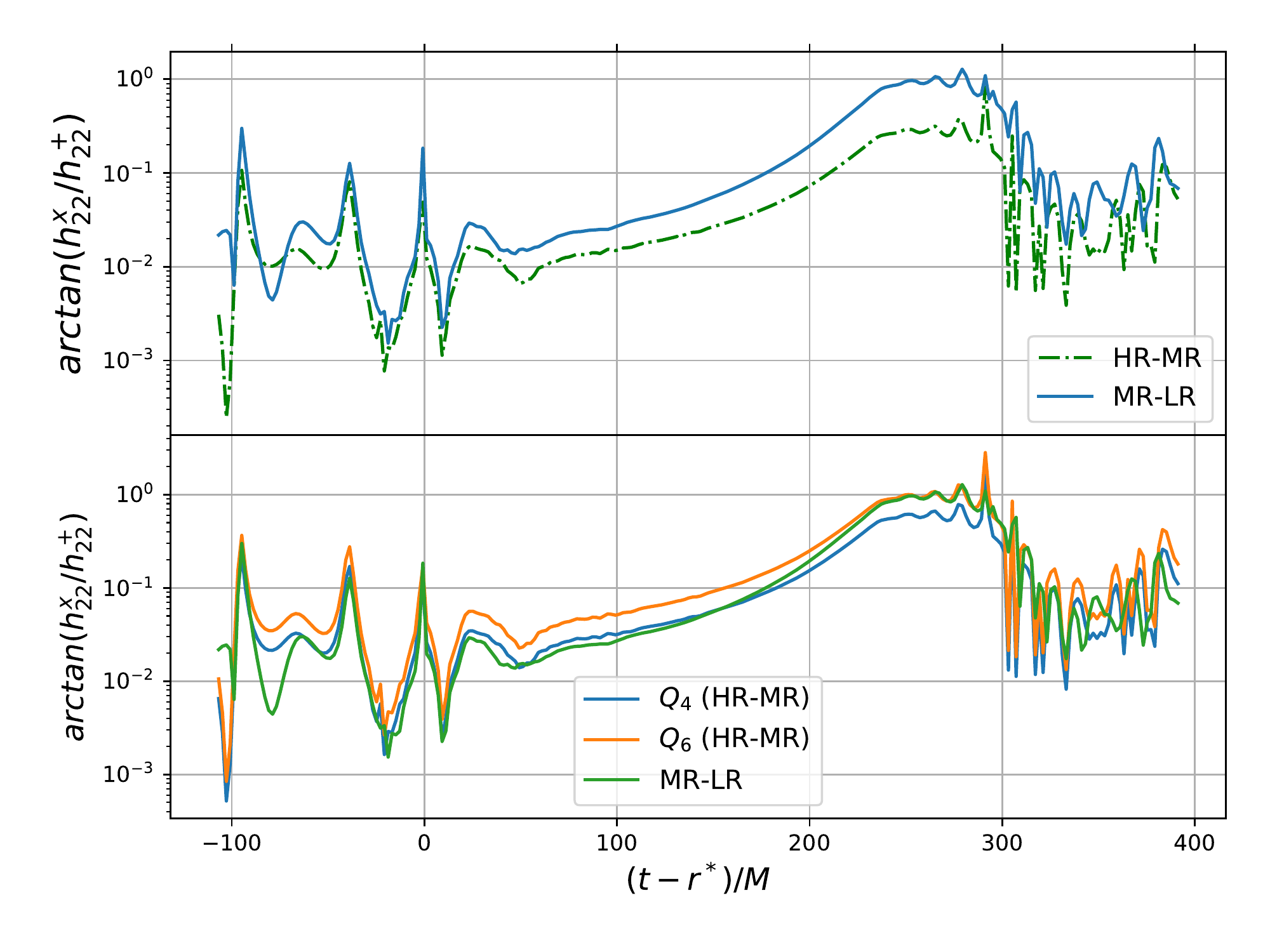}
\caption{\textit{Top}: Differences of across resolutions of the phase difference between the two polarizations of the  $(\ell,m)=(2,2)$ mode of the strain. \textit{Bottom}: Same as in the panel above but now the difference between the high and the medium resolution runs has been rescaled by the convergence factor $Q_n\equiv\frac{h_{LR}^n-h_{MR}^n}{h_{MR}^n-h_{HR}^n}$ assuming fourth (blue) and sixth (orange) order convergence.}
\label{fig:convergence}
\end{figure}

In this section we present the convergence tests for the phase difference between the two polarizations of the $(2,2)$ mode of the strain, since this is a rather standard test. More thorough tests will be presented in \cite{long_paper}. To carry out the tests, we have considered the example of a binary black holes merger in the $4\partial \text{ST}$ theory presented in the main text with the coupling constants $\lambda^{GB}/M^2=0.05$ and $g_2/M^2=1$ respectively. Therefore, this example is not in the weakly coupled regime and hence showing convergence in this case is non-trivial.

We consider three runs on a computational domain of fixed size $\Delta=512M$ and three different resolutions on the coarsest level with grid spacings $h_{LR}=\Delta/96$, $h_{MR}=\Delta/128$ and $h_{HR}=\Delta/160$ respectively. For each of these three runs, we added the same number of refinement levels, namely 8 (so the total number of levels is 9). The results presented in the main text were obtained with the medium resolution.

In Fig. \ref{fig:convergence} we show the differences between resolutions the phase difference between the two polarizations of the $(\ell,m)=(2,2)$ mode of the strain. This figure shows that during the inspiral and merger phases of the binary, the convergence order is around four and it  increases to six after the merger, which is consistent with the order of the finite difference stencils used. This mild overconvergence that \texttt{GRChombo} exhibits in the phase was already observed in the detailed studies that \cite{Radia:2021smk} carried out. The results of convergence analysis presented in this section indicate that our simulations are stable and in the convergent regime.

\bibliographystyle{apsrev4-2}
\bibliography{ref}

\end{document}